\documentclass[10pt]{article}

\usepackage{latexsym, amsmath, amssymb, amsthm}

\newtheorem{theorem}{\bf Theorem}
\newtheorem{lemma}[theorem]{\bf Lemma}

\newtheorem{proposition}[theorem]{\bf Proposition}

\newtheorem{remark}[theorem]{\bf Remark}

\newtheorem{conjecture}[theorem]{\bf Conjecture}
\newtheorem{definition}[theorem]{\bf Definition}

\def\dim{{\rm dim}}

\def\F{{\mathbb F}}
\def\Z{{\mathbb Z}}
\def\Q{{\mathbb Q}}

\def\Aalg{{\cal A}}

\def\Salg{{\cal S}}
\def\forevery{{\mbox{ for every }}}

\def\Part{{\cal P}}
\def\Coll{{\Pi}}
\def\Symm{{\rm Symm}}
\def\Orb#1#2{{\rm Orb}(#1,#2)}
\def\Supp{{\rm Supp}}
\def\Compli{{\rm I}^\perp}
\def\comm#1{}

\begin{document}
\title{\bf Schemes for Deterministic Polynomial Factoring}
\author{G\'abor Ivanyos
\footnote{Computer and Automation Research Institute
of the Hungarian Academy of Sciences,
L\'agym\'anyosi u. 11,
1111 Budapest, Hungary. 
E-mail: {\tt Gabor.Ivanyos@sztaki.hu}
}
\and
Marek Karpinski
\footnote{Department of Computer Science, University of Bonn, 53117 Bonn, Germany.
E-mail: {\tt marek@cs.uni-bonn.de}
}
\and 
Nitin Saxena
\footnote{Hausdorff Center for Mathematics, Endenicher Allee 60, 53115 Bonn, Germany.
E-mail: {\tt ns@hcm.uni-bonn.de}
}
}

\maketitle

\begin{abstract}
In this work we relate the deterministic complexity of factoring polynomials (over finite 
fields) to certain combinatorial objects we call $m$-schemes. We extend the known conditional 
deterministic subexponential time polynomial factoring algorithm for finite fields to 
get an underlying $m$-scheme. We demonstrate how the properties of $m$-schemes 
relate to improvements in the deterministic complexity of factoring polynomials 
over finite fields assuming the generalized Riemann Hypothesis (GRH). In 
particular, we give the first deterministic polynomial time algorithm (assuming GRH) to 
find a nontrivial factor of a polynomial of prime degree $n$ where $(n-1)$ is a 
smooth number.
\\\\
Keywords: Polynomials, Factoring, Deterministic, Schemes, GRH.
\end{abstract}

\section{Introduction}

We consider the classical problem of finding a nontrivial factor of a given polynomial 
over a finite field. This problem has various randomized polynomial time 
algorithms -- Berlekamp \cite{ber}, Cantor and Zassenhaus \cite{cantor-z}, von 
zur Gathen and Shoup \cite{gathen-shoup}, Kaltofen and Shoup \cite{kaltofen-shoup} 
-- but its deterministic complexity is a longstanding open problem. In this paper we study the 
deterministic complexity of the problem assuming the generalized Riemann Hypothesis (GRH). The assumption of GRH 
in this paper is needed only to find primitive $r$-th nonresidues in a finite field $\F_q$ 
which are in turn used to find a root $x$ (if it exists in $\F_q$) of ``special'' polynomials: 
$x^r-a$ over $\F_q$ (see \cite{ev-solvable}).

Assuming GRH, there are many deterministic factoring algorithms known but all of them are
exponential-time except on special instances. R\'onyai \cite{ro4} showed under GRH that any
polynomial $f(x)\in\Z[x]$, such that $\Q[x]/(f)$ is a Galois extension, can be factored modulo $p$ in 
deterministic polynomial time except for finitely many primes $p$.  
R\'onyai's result generalizes 
previous results by Huang \cite{huang}, Evdokimov \cite{ev-solvable} and Adleman, Manders and 
Miller \cite{amm}. Over special finite fields, Bach, von zur Gathen and Lenstra \cite{bgl} showed 
that polynomials over finite fields of characteristic $p$ can be factored in deterministic polynomial 
time if $\phi_k(p)$ is smooth for some integer $k$, where $\phi_k(x)$ is the $k$-th cyclotomic
polynomial. This result generalizes the previous works of R\'onyai \cite{ro2}, Mignotte and Schnorr 
\cite{mignotte}, von zur Gathen \cite{gathen}, Camion \cite{camion} and Moenck \cite{moenck}.

The line of research that we extend in this paper was started by R\'onyai \cite{ro1}. There it was 
shown how to use GRH to find a nontrivial factor of a polynomial $f(x)$, where the degree $n$ of $f(x)$ 
has a small prime factor, in deterministic polynomial time. The basic idea of \cite{ro1}, in the case when
$n$ is even, was to go to a ring extension $\Aalg^{(2)}:=\F_q[x_1,x_2]/(f(x_1),f_2(x_1,x_2))$ of
$\Aalg^{(1)}:=\F_q[x_1]/(f(x_1))$, where $f_2(x_1,x_2):=\frac{f(x_2)}{x_2-x_1}$, and then use the symmetry 
of $\Aalg^{(2)}$ to decompose $\Aalg^{(2)}$ under GRH. A decomposition of $\Aalg^{(2)}$ gives us a 
nontrivial factor of $f(x)$ since $n$ is even. \cite{ro1} showed that this basic idea can be extended to
the case when a prime $r|n$ but then the deterministic algorithm finds a nontrivial factor of $f(x)$ in time
$poly(\log q, n^r)$. The $n^r$ dependence appears in the complexity estimate because this is roughly the 
dimension of the algebras, like:
\begin{equation}\label{eqn-d-to-r}
\F_q[x_1,\ldots,x_r]/(f(x_1),\ldots,f_r(x_1,\ldots,x_r))
\end{equation}
in which the algorithm does computation.
Naively, it would seem that this algorithm will take time $poly(\log q, n^n)$ in the worst 
case (for example when $n$ is a prime). But Evdokimov \cite{ev} showed that R\'onyai's algorithm 
can be modified such that it is enough to work with algebras like (\ref{eqn-d-to-r}) with 
$r=\log n$, thus, polynomial factoring can be done deterministically in time $poly(\log q, n^{\log n})$
under GRH.

We extend Evdokimov's algorithm and show that our algorithm has an underlying natural combinatorial
structure that we call an $m$-scheme (a generalization 
of superschemes introduced by Smith \cite{smith}). An {\em $m$-scheme on $n$ points}
is, roughly speaking, a partition $\Part$ of the set $[n]^m$, where $[n]$ denotes the set $\{1,\ldots,n\}$:
$$[n]^m = \cup_{P\in\Part} P$$ 
that satisfies certain ``natural'' properties (defined in Section \ref{sec-m-scheme}). There is an 
abundance of examples of $m$-schemes in algebraic combinatorics: 
\begin{itemize}
\item a regular graph on $n$ vertices is an example of a $2$-scheme on $n$ points,
\item a strongly regular graph on $n$ vertices is an example of a $3$-scheme on $n$ points,
\item an association scheme (see \cite{book-zies}) gives rise to a $3$-scheme and vice-versa.
         See Section \ref{sec-coh-config} for these kind of examples.
\item $n$-schemes on $n$ points {\em always} arise from groups. See Section \ref{sec-orb-sch} for 
      constructing them from groups and \cite{smith} for the converse. This important example suggests 
      that $m$-schemes can be considered as a generalization of finite groups.  
\item curiously enough, $m$-schemes on $n$ points also appear when the $(m-1)$-dimensional 
      Weisfeiler-Lehman method for graph isomorphism is applied to a graph on $n$ vertices, see \cite{cfi}.
\end{itemize}
The $m$-schemes that appear in our polynomial factoring algorithm possess a special structure and we
believe that their properties can be exploited to get a deterministic and efficient polynomial
factoring algorithm (under GRH). We demonstrate that this belief infact works in several cases.

It is a standard result that to solve polynomial factoring it is enough to factor polynomials that split
completely over prime fields (see Berlekamp \cite{ber,ber70} and Zassenhaus \cite{zassen}). Thus,
we will assume in this paper that the input polynomial $f(x)$ of degree $n$ has $n$ distinct roots in 
$\F_p$ for some prime $p$. Our algorithm for factoring $f(x)$ constructs an $r$-scheme on the 
$n$ roots while working in the algebra of Equation (\ref{eqn-d-to-r}), over a suitable $\F_q\supseteq\F_p$. 
We give several results in this 
work showing how to utilise the properties of these underlying $r$-schemes to efficiently find a 
nontrivial factor of $f(x)$.

The paper is organized as follows. We formally define $m$-schemes in Section 2 and exhibit two 
important examples. In Section 3 we introduce our framework of the tensor powers 
$\Aalg^{\otimes m}$ of the algebra $\Aalg:=\F_p[x]/(f(x))$ and present our algorithm that constructs 
an underlying $m$-scheme, on the $n$ roots of $f(x)$, while working in $\Aalg^{\otimes m}$. In Section 4 we
show how to interpret Evdokimov's subexponential algorithm in our framework of $m$-schemes and give a 
conjecture about the structure of $m$-schemes which if true would make
our algorithm deterministic polynomial time under GRH. We also prove the conjecture in the important
example of $m$-schemes arising from groups. In Section 5 we show that our 
framework of $m$-schemes finds a nontrivial factor of $f(x)$ in deterministic polynomial time under GRH if $n$ is a 
prime and $(n-1)$ is smooth. In Section 6 we show that the {\em levels} $r$ (as in Equation (\ref{eqn-d-to-r}))
in Evdokimov's algorithm can be reduced to $\frac{\log n}{1.5}$ using properties of $m$-schemes. 
In Section 7 we introduce a concept of {\em primitivity} in $m$-schemes, inspired 
from the connectivity of graphs, and give some hints how it could improve the factoring algorithm. 

\section{Introducing $m$-schemes}\label{sec-m-scheme}

In this section we define special partitions of the set $[n]^m$ that we call $m$-schemes on $n$
points. These combinatorial objects are closely related to superschemes which were first defined 
by \cite{smith}.

\subsection{Basic definitions}

Let $V=\{v_1,\ldots,v_n\}$ be a set of $n$ distinct elements. For $1\leq s\leq n$, define
the set of $s$-tuples:
$$V^{(s)}:=\{(v_{i_1},\ldots,v_{i_s})\in V^s\ |\
v_{i_1},\ldots,v_{i_s}\text{ are }s\text{ distinct elements of }V\}.$$
If $s>1$ there are $s$ projections 
$\pi^s_1,\ldots,\pi^s_s: V^{(s)}\rightarrow V^{(s-1)}$ given as:
$$\pi^s_i:(v_1,\ldots,v_{i-1},v_i,v_{i+1},\ldots,v_s)
\mapsto 
(v_1,\ldots,v_{i-1},v_{i+1},\ldots,v_s).
$$
The symmetric group on $s$ elements $\Symm_s$ acts on $V^{(s)}$ in a natural way
by permuting the coordinates of the $s$-tuples. To be more accurate, the action 
is the following: for $\sigma\in \Symm_s$, 
$$(v_1,\ldots,v_i,\ldots,v_s)^\sigma=
(v_{1^\sigma},\ldots,v_{i^\sigma},\ldots,v_{s^\sigma}).$$

For $1\le m\leq n$ an {\em $m$-collection} on $V$ is a collection $\Coll$
of partitions 
$\Part_1,\Part_2,\ldots,\Part_m$ of $V=V^{(1)},V^{(2)},\ldots,V^{(m)}$
respectively. For $1\leq s\leq m$ we denote by $\equiv_{\Part_s}$ the
equivalence relation on $V^{(s)}$  corresponding to the partition $\Part_s$.
We call the equivalence classes of $\equiv_{\Part_s}$ {\em colors at level }$s$. 

We define below some natural properties of collections that are relevant to us. 
Let $\Coll=\{\Part_1,\Part_2,\ldots,$ $\Part_m\}$ be an $m$-collection
on $V$. 

{\bf Compatibility: }
We say that $\Coll$ is {\em compatible} at level $1<s\leq m$ if
${\bar u},{\bar v}\in P\in \Part_s$
implies that for every $1\leq i\leq s$ there exists $Q\in \Part_{s-1}$ 
such that $\pi_i^s({\bar u}),\pi_i^s({\bar v})\in Q$. In other words,
if two tuples (at level $s$) have the same color then for every projection
the projected tuples (at level $s-1)$ have the same color as well. 
It follows that for a class $P\in\Part_s$, the sets
$\pi_i^s(P):=\{\pi_i^s({\bar v})|{\bar v}\in P\}$, for all $i\in[s]$, are colors in 
$\Part_{s-1}$.

{\bf Regularity: }
We say that $\Pi$ is {\em regular} at level $1<s\leq m$ if
${\bar u},{\bar v}\in Q\in \Part_{s-1}$
implies that for every $1\leq i\leq s$ and for every $P\in \Part_s$,
$$\#\{{\bar u}^\prime\in P\ |\ \pi_i^s({\bar u}^\prime)={\bar u}\}=
\#\{{\bar v}^\prime\in P\ |\ \pi_i^s({\bar v}^\prime)={\bar v}\}$$
We call the tuples in $P\cap(\pi_i^s)^{-1}({\bar u})$ as {\em $\pi_i^s$-fibers}
of ${\bar u}$ in $P$. So regularity, in other words, means that the cardinalities of the fibers 
above a tuple depend only on the color of the tuple.

The above two properties motivate the definition of the {\em subdegree of a color $P$ over a
color $Q$} as $\frac{\#P}{\#Q}$ when $\Pi$ is compatible and regular at level $s$ and 
$\pi_i^s(P)=Q$ for some $i$.

{\bf Invariance: }
An $m$-collection is {\em invariant} at level $1<s\leq m$ if
for every $P\in\Part_s$, and $\sigma\in\Symm_s$ we have:
$$P^\sigma := \{{\bar v}^\sigma|{\bar v}\in P\}\in \Part_s.$$
In other words, the partitions $\Part_1,\ldots,\Part_m$ are
invariant under the action of the corresponding symmetric group.

{\bf Homogeneity: }
We say that the $m$-collection $\Coll$ is {\em homogeneous} if $|\Part_1|=1$.

{\bf Symmetry: }
We say that an $m$-collection $\Coll$ is {\em symmetric}
at level $s$ if for every $P\in\Part_s$ and $\sigma\in\Symm_s$, we have $P^\sigma=P$.

{\bf Antisymmetry: }
We say that an $m$-collection $\Coll$
is {\em antisymmetric} at level $s$ if 
for every $P\in\Part_s$ and $1\not=\sigma\in\Symm_s$,
we have $P^\sigma\not=P$.

\begin{definition}
An $m$-collection is called compatible, regular, invariant,
symmetric, or antisymmetric
if it is at every level $1<s\leq m$ compatible, regular, invariant,
symmetric, or antisymmetric respectively. 

An $m$-collection is called an {\em $m$-scheme} if it is compatible, regular and invariant. 
\end{definition}

We should remark that the $m$-schemes that appear in our factoring algorithm
are homogeneous and antisymmetric as well. Let us now see some easily describable 
examples of $m$-schemes.

\subsection{Example: $3$-schemes from coherent configurations}\label{sec-coh-config}

Coherent configurations are standard combinatorial objects that have strongly regular graphs as 
examples (see \cite{book-cameron}). Recall that a coherent configuration is just a $2$-scheme
$\{\Part_1, \Part_2\}$ that also has a composition property:

{\bf Composition: }
For any $P_i, P_j, P_k\in\Part_2$ and an $(\alpha,\beta)\in P_k$ the number:
$$\#\{\gamma\in V\mid (\alpha,\gamma)\in P_i \text{ and }(\gamma,\beta)\in P_j\}$$
is independent of which tuple $(\alpha,\beta)$ in $P_k$ we chose.
In other words, the relations $P_i$ and $P_j$ can be ``composed'' to get a bigger relation that is
just a ``linear combination'' of the relations in $\Part_2$.

In the literature a homogeneous coherent configuration is usually called an {\em association
scheme}. In this paper we do not enforce symmetricity or antisymmetricity in the definition of
an association scheme. Coherent configurations and $3$-schemes are similar notions. 

{}From a coherent configuration $\{\Part_1,\Part_2\}$ we can define a partition $\Part_3$ on the 
triples such that for any two triples $(u_1,u_2,u_3)$ and $(v_1,v_2,v_3)$ we have: \\
$(u_1,u_2,u_3)\equiv_{\Part_3}(v_1,v_2,v_3)$ if and only if
$(u_1,u_2)\equiv_{\Part_2}(v_1,v_2)$,
$(u_1,u_3)\equiv_{\Part_2}(v_1,v_3)$,
$(u_2,u_3)\equiv_{\Part_2}(v_2,v_3)$.\\
It follows that for $P\in\Part_3$, the cardinality $\#\{u_3\in V|(u_1,u_2,u_3)\in P\}$ 
of the $\pi_3^3$-fibers of $(u_1,u_2)$ in $P$ is
exactly  $\#\{u_3\in V|(u_1,u_3)\in \pi_2^3(P)\mbox{ and } (u_2,u_3)
\in \pi_1^3(P)\}$ and thus regularity at level $3$ is equivalent to the composition property of
$\{\Part_1,\Part_2\}$. It is easy to show that $\{\Part_1,\Part_2,\Part_3\}$ also satisfies
compatibility and invariance, thus, it is a $3$-scheme.

Similarly, a converse can be shown:
\begin{lemma}\label{lem-3-sch-assoc-sch}
If $\ \Pi=\{\Part_1,\Part_2,\Part_3\}$ is a homogeneous $3$-scheme then
$\{\Part_1,\Part_2\}$ is an association scheme.
\end{lemma}
\begin{proof}
By the hypothesis we already have that $\{\Part_1,\Part_2\}$ is a homogeneous
$2$-scheme. Thus, we only need to show the composition property.
Let $P_i,P_j,P_k\in\Part_2$ and let $(\alpha,\beta)\in P_k$. Then by compatibility at level $3$ there exists a 
subset $\Salg\subseteq\Part_3$ such that the set:
$$\{\gamma\in V\mid (\alpha,\gamma)\in P_i, (\gamma,\beta)\in P_j\}$$
can be partitioned as:
$$\sqcup_{P\in \Salg}\{\gamma\in V\mid (\alpha,\gamma)\in P_i, (\gamma,\beta)\in P_j, 
(\alpha, \gamma, \beta)\in P\}$$
which again by the compatibility of $\Pi$ at level $3$ is:
$$\sqcup_{P\in \Salg}\{\gamma\in V\mid (\alpha, \gamma, \beta)\in P\}$$
now by the regularity of $\Pi$ at level $3$ the size of the above sets is simply 
$\frac{\#P}{\#P_k}$ which is independent of the choice of $(\alpha,\beta)$.
Thus, $\{\Part_1,\Part_2\}$ has the composition property.
\end{proof}

\subsection{Example: orbit schemes}\label{sec-orb-sch}

Permutation groups provide a host of examples (see \cite{smith}).
Let $G\leq\Symm_V$ be a permutation group. The {\em orbits} of
$G$ on the $s$-tuples ($1\le s\le m\le n$) give an $m$-scheme. More formally, define
the partition $\Part_s$ as: for any two $s$-tuples $(u_1,\ldots,u_s)$ and $(v_1,\ldots,v_s)$ in 
$V^{(s)}$, $(u_1,\ldots,u_s)\equiv_{\Part_s}(v_1,\ldots,v_s)$ iff $\exists\sigma\in G$, 
$(\sigma(u_1),\ldots,\sigma(u_s))=$ $(v_1,\ldots,v_s)$. It is easy to see that these partitions
naturally satisfy compatibility, regularity and invariance properties and hence form an 
$m$-scheme. We call $m$-schemes arising in this way {\em orbit $m$-schemes}.

The orbit scheme is homogeneous if and only if $G$ is transitive.
Furthermore, assume that $G$ is transitive and for some integer
$m< n$, $\gcd(m!,|G|)=1$. Then the corresponding orbit $m$-scheme is
a homogeneous antisymmetric $m$-scheme.
Our attention to this class of examples has been drawn by D.~Pasechnik. 

At the moment, we are
not aware of any other examples of  homogeneous antisymmetric $m$-schemes with 
$m\rightarrow\infty$. The homogeneous antisymmetric $m$-schemes are the ones that arise 
in our factoring algorithm and we do believe that their parameters satisfy more stringent 
conditions than the general $m$-schemes. For a conjecture along these lines see Section 
\ref{sec-conj}.

\subsection{Difference between various notions of schemes}

The term {\em schemes} arises in the mathematical literature in many contexts. Our $m$-schemes 
should not be confused with the notion of {\em schemes} in algebraic geometry. However, our 
$m$-schemes are closely related to {\em association schemes}, {\em superschemes} (Smith 
\cite{smith}) and {\em height $t$ presuperschemes} (Wojdy\l o \cite{wojd}). Smith's superschemes 
are $m$-schemes that also satisfy a suitable higher dimensional generalization of the composition 
property. It is not difficult to see that a superscheme on $n$ points is just a $n$-scheme on $n$ points.
Wojdy\l o's height $t$ presuperscheme consists of the bottom $t$ levels of a superscheme.
In particular, a level $0$ presuperscheme is just an association scheme.
It can be shown that a height $t$ presuperscheme on $n$-points 
consists just of the first $(t+2)$ levels of a $(t+3)$-scheme on $n$ points.

\section{Decomposition of tensor powers of algebras}\label{sec-tensors}

In this section we describe our polynomial factoring algorithm and simultaneously
show how $m$-schemes appear in the algorithm. Recall that in the input we are given a
polynomial $f(x)\in\F_p$ of degree $n$ having distinct roots $\alpha_1,\ldots,\alpha_n$ in $\F_p$. 
For any extension field $k$ of $\F_p$ we have the 
natural associated algebra $\Aalg:=k[X]/(f(X))$. Note that $\Aalg$ is a completely split 
semisimple $n$-dimensional algebra over the field $k$, i.e. $\Aalg$ is isomorphic to 
$k^n$ the direct sum of $n$ copies of the one-dimensional $k$-algebra $k$.
We interpret $\Aalg$ as the set of functions: 
$$V:=\{\alpha_1,\ldots,\alpha_n\} \rightarrow k$$
equipped with the pointwise operations.
Algorithmically, we have $\Aalg$ by structure constants with respect to some
basis $b_1,\ldots,b_n$ (for example, $1,X,\ldots,X^{n-1}$) and the problem of factoring 
$f(X)$ completely can be viewed as finding an explicit isomorphism from 
$\Aalg$ to $k^n$. 

{\em How do the factors of $f(X)$ appear in $\Aalg$?} They appear as {\em zero divisors} 
in $\Aalg$. Recall that a zero divisor is a nonzero element $z(X)\in\Aalg$
such that $y(X)z(X)=0$ for some nonzero element $y(X)\in \Aalg$. This means that
$f(X)|y(X)\cdot z(X)$ which implies (by the nonzeroness of $y$ and $z$) $gcd(f(X),z(X))$ 
factors $f(X)$ nontrivially. As gcd of polynomials can
be computed by the deterministic polynomial time Euclidean algorithm,
we infer that finding a zero divisor in the factor 
algebra $k[X]/(f(X))$ is -- up to polynomial time deterministic
reductions -- equivalent to finding a nontrivial divisor 
of $f(X)$. Furthermore, computing an explicit isomorphism with $k^n$
is equivalent to factoring $f(X)$ completely.

{\em How are the ideals of $\Aalg$ related to the roots of $f(x)$?} Let $I$ be an ideal of 
$\Aalg$. The {\em support of $I$}, $\Supp(I)$ is defined as
$$\Supp(I):=V\setminus\{v\in V\ |\ a(v)=0 \forevery a\in I\}$$
Conversely, for $U\subseteq V$, the ideal ${\rm I}(U)$ is defined as:
$${\rm I}(U):=\{b\in \Aalg\ |\ b(u)=0 \forevery u\in U\}$$
and $\Compli(U)$ is the {\em annihilator} of ${\rm I}(U)$:
$$\Compli(U):=\{a\in \Aalg\ |\ ab=0 \forevery b\in {\rm I}(U)\}.$$
It can be easily seen that $\Supp$ is an inclusion preserving
 bijection from the ideals of $\Aalg$ to the subsets
of $V$ with inverse map $\Compli$.
In view of this correspondence, partial decompositions of 
$\Aalg$ into sums of pairwise orthogonal 
ideals correspond to partitions of the set $V$. Let us formulate the above discussion in a 
lemma.
\begin{lemma}\label{lem-m-is-1}
If $I_1,\ldots,I_t$ are pairwise orthogonal ideals of $\Aalg$ such that 
$\Aalg=I_1+\cdots+I_t$ then $V=\Supp(I_1)\sqcup\cdots\sqcup\Supp(I_t)$.
\end{lemma}

We now move up to the tensor powers of $\Aalg$ and there we show a way of
getting the partitions of $V^{(m)}$. For $m\in[n]$, let $\Aalg^{\otimes m}$ 
denote the $m$th tensor power of $\Aalg$.
$\Aalg^{\otimes m}$ is also a completely split semisimple algebra; it is isomorphic
to $k^{n^m}$. We again interpret it as the algebra of functions from $V^m$ to 
$k$. Note that in this interpretation the rank $1$ tensor element 
$h_1\otimes\cdots\otimes h_m$ corresponds to a function 
$V^m\rightarrow k$ that maps $(x_1,\ldots,x_m)\mapsto h_1(x_1)\cdots h_m(x_m)$ .

The {\em essential part} $\Aalg^{(m)}$ of $\Aalg^{\otimes m}$
is the ideal consisting of the functions which vanish
on all the $m$-tuples $(v_1,\ldots,v_m)$ with
$v_i=v_j$ for some $i\neq j$. Then $\Aalg^{(m)}$
can be interpreted as the algebra of functions $V^{(m)}\rightarrow k$. We show below that a basis for $\Aalg^{(m)}$ can be 
computed easily and then this is the algebra where our factoring algorithm does computations.
\begin{lemma}\label{lem-compute-A-to-m}
Given $f(X)$, a polynomial of degree $n$ having $n$ distinct roots in $\F_p$,
a basis for $\Aalg^{(m)}=\left(k[X]/(f(X))\right)^{(m)}$ over $k\supseteq\F_p$ can be computed 
by a deterministic algorithm in time $poly(\log p, n^m)$. 
\end{lemma}
\begin{proof}
To see this, consider embeddings $\mu_i$ of $\Aalg$ into
$\Aalg^{\otimes m}$ ($i=1,\ldots,m$) given as $\mu_i(a)=
1\otimes\ldots\otimes 1 \otimes a \otimes 1 \otimes \ldots\otimes 1$ where $a$ is 
of course in the $i$-th place. In the interpretation as functions,
$\mu_i(\Aalg)$ correspond to the functions
on $V^m$ which depend only on the $i$th element in the tuples.
Observe that the set, for $1\le i< j\le m$:
$$\Delta^{m}_{i,j}=\{b\in \Aalg^{\otimes m}\ |\ (\mu_i(a)-\mu_j(a))b=0
\forevery a\in \Aalg\}$$
is the ideal of $\Aalg^{\otimes m}$ consisting of the functions which are zero on 
every tuple $(v_1,\ldots,v_m)$ with $v_i\neq v_j$. Given a basis for
$\Aalg$, a basis for $\Delta^m_{i,j}$ can be computed
by solving a system of linear equations in time
polynomial in the dimension of $\Aalg^{\otimes m}$ (over $k$) which is $n^m$. 
Finally, notice that $\Aalg^{(m)}$ can be computed as well since it is the annihilator of
$\sum_{1\le i<j\le m}\Delta^{m}_{i,j}$.
\end{proof}

\begin{remark}
The algebras $\Aalg^{(m)}$ which we are now going to work with have a simple
explicit description, for example, $\Aalg^{(1)}$ is ofcourse $k[X_1]/(f(X_1))$ and $\Aalg^{(2)}$ is nothing but 
$k[X_1,X_2]/(f(X_1),$ $f_2(X_1,X_2))$ where $f_2(X_1,X_2)$ is a polynomial in $\Aalg^{(1)}[X_2]$ 
defined as $\frac{f(X_2)}{X_2-X_1}$. Similarly, we can write down an expression for $\Aalg^{(m)}$ inductively.
\end{remark}

Like the case of $m=1$, ideals and partial decompositions of $\Aalg^{(m)}$ into 
pairwise orthogonal ideals correspond to subsets and 
partitions of the set $V^{(m)}$ respectively. If $I$ is an ideal of $\Aalg^{(m)}$ then we again 
define the {\em support of $I$}, $\Supp(I)$ as:
$$\Supp(I):=V^{(m)}\setminus\{\bar{v}\in V^{(m)}\ |\ a(\bar{v})=0 \forevery a\in I\}$$
Lemma \ref{lem-m-is-1} generalizes to:
\begin{lemma}\label{lem-ideal-set}
For any $s\le n$, if $I_{s,1},\ldots,I_{s,t_s}$ are pairwise orthogonal ideals of 
$\Aalg^{(s)}$ such that $\Aalg^{(s)}=I_{s,1}+\cdots+I_{s,t_s}$ then 
$V^{(s)}=\Supp(I_{s,1})\sqcup\cdots\sqcup\Supp(I_{s,t_s})$.
\end{lemma}
Now we will describe our polynomial factoring algorithm that produces $m$-schemes.

\subsection*{Algorithm Description}\label{sec-poly-to-sch}
{\bf Input:} a degree $n$ polynomial $f(x)$ having $n$ distinct roots in $\F_p$. Given $1<m\le n$
we can wlog assume that we also have the smallest field extension $k\supseteq\F_p$ having $s$-th nonresidues 
for all $s\in[m]$ (computing $k$ will take $poly(\log p, m^m)$ time under GRH).
\\\\
{\bf Output:} a nontrivial factor of $f(x)$ or a homogeneous, antisymmetric $m$-scheme 
on the $n$ points: $V:=\{\alpha\in\F_p|f(\alpha)=0\}$.
\\\\
{\bf Algorithm overview:}
\\\\
We define $\Aalg^{(1)}=\Aalg=k[x]/(f(x))$ and compute $\Aalg^{(s)}$, for all $s\in[m]$, in time
$poly(\log p, n^m)$ (by Lemma \ref{lem-compute-A-to-m}). 

Now observe that $Aut_k(\Aalg^{(s)})$ contains $\Symm_s$. To see this, just
note that there is an action of $\Symm_s$ on $\Aalg^{(s)}$ as 
a group of algebra automorphism, for $\sigma\in \Symm_s$ this action
is the linear extension of:
$$(b_{i_1}\otimes\cdots\otimes b_{i_s})^\sigma=
b_{i_{1^\sigma}}\otimes\cdots\otimes b_{i_{s^\sigma}}.$$
This knowledge of explicit automorphisms of $\Aalg^{(s)}$ 
can be exploited to efficiently decompose these algebras under GRH (see Theorem 2.3 in \cite{ro4}). 
Thus, for all $1<s\le m$ we can compute mutually orthogonal $t_s\ge2$ ideals $I_{s,i}$ of $\Aalg^{(s)}$, such 
that:
$$\Aalg^{(s)}=I_{s,1}+\ldots+I_{s,t_s}$$
By Lemma \ref{lem-ideal-set}, the above decomposition induces partitions $\Part_s$ for 
all $1<s\le m$ such that:
$$\Part_s:\ \ V^{(s)}=\Supp(I_{s,1})\sqcup\cdots\sqcup\Supp(I_{s,t_s})$$
Thus, together with $\Part_1:=\{V\}$ we have an $m$-collection $\Pi:=(\Part_1,\ldots,\Part_m)$ on 
the set $V$.

Now we will show how to refine this $m$-collection to an $m$-scheme using algebraic
operations on the ideals $I_{s,i}$ of $\Aalg^{(s)}$. To do that, we first need a tool
to relate lower level ideals $I_{s-1,i}$ to higher level ideals $I_{s,i^\prime}$.
For every $1<s\leq m$, we have $s$ embeddings
$\iota_j^s:\Aalg^{\otimes (s-1)}\rightarrow\Aalg^{\otimes s}$ 
sending $b_{i_1}\otimes\cdots\otimes b_{i_{s-1}}$ to
$b_{i_1}\otimes\cdots\otimes b_{i_{j-1}}\otimes 1\otimes
b_{i_j}\otimes\cdots b_{i_{s-1}}$. Restricting to $\Aalg^{(s-1)}$ and
multiplying the images of $\iota_j^s$ by the identity element of $\Aalg^{(s)}$,
we obtain algebra embeddings $\Aalg^{(s-1)}\rightarrow\Aalg^{(s)}$
 denoted also by $\iota_1^s,\ldots,\iota_s^s$.
In the function interpretation, $\iota_j^s(\Aalg^{(s-1)})$ is just the set of functions
in $\Aalg^{(s)}$ which do not depend on the $j$th coordinate of tuples.

{\bf Compatibility} of the $m$-collection $\Pi$ at level $1<s\le m$ corresponds to:
for every pair of ideals $I_{s-1,i}$ and $I_{s,i^\prime}$ in 
the decomposition of $\Aalg^{(s-1)}$ and $\Aalg^{(s)}$ respectively 
and for every $j\in\{1,\ldots,s\}$, the ideal 
$\iota_j^s(I_{s-1,i})I_{s,i^\prime}$ can be assumed to be either zero or $I_{s,i^\prime}$.
Otherwise we can efficiently compute a subideal of $I_{s,i^\prime}$, hence,
refining $I_{s,i^\prime}$ and the $m$-collection $\Pi$.

{\bf Regularity} of the $m$-collection $\Pi$ at level $1<s\le m$ corresponds to:
for every pair of ideals $I_{s-1,i}$ and $I_{s,i^\prime}$ in
the decomposition of $\Aalg^{(s-1)}$ and $\Aalg^{(s)}$, respectively,
and for every $j\in\{1,\ldots,s\}$, 
$\iota_j^s(I_{s-1,i})I_{s,i^\prime}$ can be assumed to be a free module over 
$\iota_j^s(I_{s-1,i})$.
Otherwise by trying to find a free basis, we can efficiently compute a zero divisor in 
$I_{s-1,i}$, hence, refining $I_{s-1,i}$ and the $m$-collection $\Pi$.

Compatibility and regularity of $\Pi$ create a natural connection between the ideals of levels 
$(s-1)$ and $s$, for all $1<s\le m$. In the case when a pair of ideals $I_{s-1,i}$ and 
$I_{s,i^\prime}$ in the decomposition of $\Aalg^{(s-1)}$ and $\Aalg^{(s)}$ respectively 
satisfies $\iota_j^s(I_{s-1,i})I_{s,i^\prime}$ $=I_{s,i^\prime}$: $I_{s,i^\prime}$ is a free 
module over $\iota_j^s(I_{s-1,i})$ which in other words means that the elements in 
$I_{s,i^\prime}$ can be viewed as univariate polynomials with coefficients in $I_{s-1,i}$. The 
rank of the free module $I_{s,i^\prime}$ over $\iota_j^s(I_{s-1,i})$ can easily be seen to be equal 
to the subdegree of $\Supp(I_{s,i^\prime})$ over $\Supp(I_{s-1,i})$. 

{\bf Invariance} of the $m$-collection $\Pi$ at level $1<s\le m$ may be assumed, 
since if for some $\sigma\in \Symm_s$ the decomposition of
$\Aalg^{(s)}$ is not $\sigma$-invariant, then
we can find two ideals $I_{s,i}$ and $I_{s,i^\prime}$ such that 
$I_{s,i}^\sigma\cap I_{s,i^\prime}$ is neither zero nor $I_{s,i^\prime}$, thus, 
we can efficiently refine $I_{s,i^\prime}$ and the $m$-collection $\Pi$.

{\bf Homogeneity} of the $m$-collection $\Pi$ corresponds to: the algebra $\Aalg^{(1)}=\Aalg$
is not in a decomposed form.

{\bf Antisymmetricity} of the $m$-collection $\Pi$ at level $1<s\le m$ corresponds to: for any 
ideal $I_{s,i}$ at level $1<s\le m$ and for any $\sigma\in\Symm_s\setminus\{id\}$, we can
assume $I_{s,i}^\sigma\not= I_{s,i}$. Otherwise $\sigma$ is an algebra automorphism of 
$I_{s,i}$ and hence we can find its subideal efficiently under GRH by \cite{ro4}, thus, refining
$I_{s,i}$ and the $m$-collection $\Pi$.

Note that invariance and antisymmetricity at level $s$ entail $s!\mid t_s$.

By the observations above: we can keep applying ideal operations in the algebras $\Aalg^{(s)}$, 
$s\in[m]$, till either we get a nontrivial factor of $f(x)$ or the underlying $m$-collection 
$\Pi$ becomes a homogeneous, antisymmetric $m$-scheme on $n$ points. The time taken by our
algorithm is clearly $poly(\log p, n^m)$.

\begin{remark}\label{rem-no-sch}
At this point we are able to reprove Ronyai's result \cite{ro1}: under GRH,
we can deterministically find a nontrivial factor of a degree $n$ polynomial over $\F_p$ 
in time $poly(\log p,$ $n^r)$, where $r$ is the smallest prime divisor of $n$.
The proof is to algorithmically try constructing an $r$-scheme as above and show by
an easy divisibility argument that there exist
no homogeneous, antisymmetric $r$-schemes on $n$ points if $r$ is a divisor of $n$. 
This guarantees that our algorithm will be forced to find a nontrivial factor of $f(x)$.
\end{remark}

\section{$m$-schemes in Evdokimov's Algorithm}\label{sec-evdokimov}

We saw in the last section how to either find a nontrivial factor of a given $f(x)$
or construct an $m$-scheme on the $n$ roots of $f(x)$. Our aim is to analyse the 
``bad case'' of the algorithm when we get no nontrivial factor but instead we get an
antisymmetric, homogeneous $m$-scheme. Can the properties of these $m$-schemes be used to 
factor $f(x)$? 
In the rest of the paper we will try to answer that question. Here we start with an
exposition of Evdokimov's idea \cite{ev} in our framework of $m$-schemes. We show below
that \cite{ev} exploited the presence of {\em matchings} in the $m$-schemes.

\begin{definition}
A color $P\in \Part_s$, for $1<s\leq m$, in an $m$-scheme $\{\Part_1,\ldots,\Part_m\}$
is called a {\em matching} if there exist $1\leq i<j\leq s$ such that $\pi_i^s(P)=\pi_j^s(P)$
and $|\pi_i^s(P)|=|P|$. 
\end{definition}

The presence of matchings can be used to efficiently refine the underlying $m$-scheme.

\begin{lemma}\label{lem-matching-refines}
If the color $P\in\Part_s$ is a matching then under GRH we can refine the $m$-scheme $\Pi=$
$\{\Part_1,\ldots,\Part_m\}$ deterministically in time $poly(\log p, n^m)$.
\end{lemma}
\begin{proof}
Following the notation of the above definition, it is obvious that if color $P$ is a matching then
both $\pi_i^s$ and $\pi_j^s$ are bijections, therefore the map $\pi_i^s(\pi_j^s)^{-1}$ is a 
permutation of $\pi_j^s(P)$. Furthermore, this permutation is nontrivial as $P\subseteq V^{(s)}$. 
So in the corresponding orthogonal ideals decomposition of $\Aalg^{(1)},\ldots,\Aalg^{(m)}$, both 
the maps $\iota_i^s$ and $\iota_j^s$ give isomorphisms 
$I_{s-1,\ell^\prime}\rightarrow I_{s,\ell}$, where the ideals $I_{s-1,\ell^\prime}$ and $I_{s,\ell}$
correspond to $\pi_j^s(P)$ and $P$ respectively. This means that the map 
$(\iota_i^s)^{-1}\iota_j^s$ is a nontrivial automorphism of $I_{s-1,\ell^\prime}$. It follows from 
\cite{ro4} that, assuming GRH, we can obtain a proper decomposition of $I_{s-1,\ell^\prime}$ and
hence refine the $m$-scheme $\Pi$.
\end{proof}

Now we show the idea of \cite{ev} to find a matching in $\log_2 n$ levels.

\begin{lemma}\label{lem-matching-in-logn}
If the $m$-scheme $\Pi:=\{\Part_1,\ldots,\Part_m\}$ on $n$ points is antisymmetric
at the second level, $|\Part_1|<n$ and $m\geq \log_2 n$ then there is a matching
in $\{\Part_1,\ldots,\Part_m\}$.
\end{lemma}
\begin{proof}
We will give an effective way of finding a matching given such a $\Pi$.
Choose $P_1\in\Part_1$
with $d_1:=|P_1|>1$. It is clear that $Q_2=P_1^{(2)}$ is a disjoint union
of some colors in $\Part_2$. Choose a smallest color $P_2\in \Part_2$ with $P_2\subseteq Q_2$. 
By the definition of an $m$-scheme: $\pi_1^2(P_2)=\pi_2^2(P_2)=P_1$.
Also, by antisymmetry we can infer that $d_2:={\frac{|P_2|}{|P_1|}}$ $<d_1/2$. If $d_2=1$ 
then $P_2$ is a matching.

If $d_2>1$ then we proceed in the following iterative way.
Suppose that, for some $2< s< m$, we have already chosen colors
$P_1\in\Part_1,\ldots,P_{s-1}\in\Part_{s-1}$ with 
$\pi_{i-1}^i(P_i)=\pi_{i}^i(P_i)=P_{i-1}$ and
$1<d_i:={\frac{|P_i|}{|P_{i-1}|}}< d_{i-1}/2$
for every $2\le i\le s-1$. Since $d_{s-1}>1$, the set 
$Q_s=\{{\bar v}\in V^{(s)}|
\pi_{s-1}^s({\bar v})\in P_{s-1},\pi_{s}^s({\bar v})\in P_{s-1}\}$
is nonempty. Let $P_s$ be a smallest class from $\Part_s$ with
$P_s\subseteq Q_s$. Again antisymmetry implies that 
$d_s:={\frac{|P_s|}{|P_{s-1}|}}$ $<d_{s-1}/2$.
If $d_s=1$ then $P_s$ is clearly a matching. Otherwise we proceed to the level $(s+1)$
and further halve the subdegree. This procedure finds a matching
in at most $\log_2 d_1\leq \log_2 n$ rounds.
\end{proof}

From our algorithm in the last section and the above two lemmas it follows that, under GRH, 
we can completely factor $f(x)$ deterministically in
$poly(\log p,$ $n^{\log n})$ time. This is the result of Evdokimov \cite{ev}.

It might be worth noting that in the above Lemma we used antisymmetry (and even
invariance) merely at level $2$. Indeed, if a compatible and regular 
$m$-collection $\{\Part_1,\ldots,\Part_m\}$ is antisymmetric at 
level $2$ then for every $1<s\leq m$ and every $s$-element subset 
$\{v_1,\ldots,v_s\}\subseteq V\;$ we have $(v_1,\ldots,v_{s-1},v_s)\not\equiv_{\Part_s}$
$(v_1,\ldots,v_s,v_{s-1})$. (This can be seen by projecting to the last two coordinates.)

\subsection{A Conjecture about Matchings}\label{sec-conj}

Here we make a conjecture about the structure of homogeneous, antisymmetric $4$-schemes and
higher schemes. It might seem a bit unmotivated but we
show below, interestingly, that it is true in the case of orbit schemes. Note that orbit schemes 
are the only (infinite) family of $4$-schemes we currently know that are homogeneous and 
antisymmetric.

\begin{conjecture}\label{conj-matchings}
There exists a constant $m\ge 4$ such that every homogeneous,
antisymmetric $m$-scheme contains a matching.
\end{conjecture}
 
It is clear by Lemma \ref{lem-matching-refines} that a proof of this conjecture would result 
in a deterministic polynomial time algorithm for factoring polynomials over 
finite fields (under GRH).

We will now show that Conjecture \ref{conj-matchings} holds, with $m=4$, for the important 
example of orbit schemes. It is easy to see that the $2$-scheme associated to a permutation group
$G$ is antisymmetric if and only if $|G|$ is odd. Assume that $G$ 
is a nontrivial permutation group of odd order on $V=\{1,\ldots,n\}$.
Let $H$ be a subgroup minimally containing the stabilizer $G_1$ of
$G$. Let $B=\Orb{H}{1}$ be the orbit of 1 under the action of $H$. Then $H$
acts as a primitive permutation group on $B$. 
Also, by \cite{seress}, there is a base of size $s\leq 3$ of $H$. 
This is a subset $\{b_1,\ldots,b_s\}\subseteq B$
such that $H_{b_1}\cap\cdots\cap H_{b_s}=N$, where 
where $N$ is the kernel of the permutation representation
 of $H$ on $B$. We assume that this base is irredundant, in particular 
$K=H_{b_1}\cap\cdots\cap H_{b_{s-1}}>N$.
Since $K_{b_s}=N<K$ there exists $b_{s+1}\in \Orb{K}{b_s}\setminus\{b_s\}$.
In order to simplify notation, we assume $b_1=1,b_2=2,\ldots,b_{s+1}=s+1$.
The first equality $b_1=1$ can be ensured using the 
transitivity of $H$ on $B$, while the
others can be achieved by renumbering $V$. {}From $G_1<H$ 
we infer that $N=H_1\cap\cdots\cap H_t=G_1\cap\cdots\cap G_t$ holds
for every $t\in\{1,\ldots,s+1\}$. Let $P$ be the $G$-orbit
of $(1,\ldots,s+1)$. Since $(1,\ldots,s-1,s)$ and $(1,\ldots,s-1,s+1)$
are in the same orbit, we have $\pi^{s+1}_s(P)=\pi^{s+1}_{s+1}(P)$.
Also, since the $(1,\ldots,s)$ and $(1,\ldots,s,s+1)$ both have 
stabilizer $N$, the size of the orbits of both tuples coincide with
$|G:N|$.  These properties imply that $P$ is a matching.

\section{Factoring polynomials of smooth prime degree}

We saw in Section \ref{sec-poly-to-sch} how to obtain a homogeneous  
$m$-scheme on $n$ points from a given polynomial of degree $n$ and we also saw 
in Lemma \ref{lem-3-sch-assoc-sch} that a homogeneous $3$-scheme is an 
association scheme. We now use a recent interesting result of Hanaki and Uno 
\cite{hu} about the structure of association schemes, on a {\em prime} number of
points, to factor polynomials when $n$ is a smooth prime number.

\begin{theorem}
If $n>2$ is prime, $r$ is the largest prime factor of $(n-1)$ and $f(x)$ is a 
degree $n$ polynomial over $\F_p$ then we can find a 
nontrivial factor of $f(x)$ deterministically in time $poly(\log p, n^r)$
under GRH.
\end{theorem}
\begin{proof}
Wlog we can assume that $f(x)$ has $n$ distinct roots ($\alpha_i$'s) in $\F_p$. {}From Section
\ref{sec-poly-to-sch} we can again assume that we have constructed a homogeneous 
antisymmetric $(r+1)$-scheme on $n$ points: $(\Part_1,\ldots,\Part_{r+1})$. Now 
from Lemma \ref{lem-3-sch-assoc-sch} we know that $(\Part_1,\Part_2)$ is an
antisymmetric association scheme. {}From \cite{hu}: $\exists d|(n-1)$, $\forall P\in\Part_2$, $\#P=dn$. 
If $d=1$ then we have matchings in $\Part_2$ and hence by Lemma 
\ref{lem-matching-refines} we can find a nontrivial factor of $f(x)$.

On the other hand if $d>1$ then the colors in $(\Part_2,\ldots,\Part_{r+1})$ naturally
induce homogeneous antisymmetric $r$-schemes on $d$ points (for example, restrict the partitions
to tuples that have $\alpha_1$ in the first coordinate). As $d$ has a prime divisor 
which is at most $r$ there do not exist such schemes by Remark \ref{rem-no-sch}.

The time complexity follows from our algorithm overview.
\end{proof}

\section{Reducing the number of levels in Evdokimov's algorithm}\label{sec-n-by-8}

We saw in Lemma \ref{lem-matching-in-logn} that a homogeneous $m$-scheme on $n$ points that 
is antisymmetric at level $2$ has a matching below the $\lceil\log_2 n\rceil$-th level. Recall
from Section \ref{sec-poly-to-sch} that from a polynomial we can construct an $m$-scheme
that is antisymmetric at every level $>1$ and not just at level $2$. Are we then 
guaranteed to get a matching at a level less than $\log n$? We conjecture that there 
should be a matching at a much smaller level as intuitively antisymmetricity reduces the 
subdegrees of the colors but we could prove only a constant fraction of $\log n$
upper bound on the number of levels. First we prove a lemma:

\begin{lemma}\label{lem-n-by-8}
Let $\Pi=(\Part_1,\ldots,\Part_4)$ be a homogeneous, antisymmetric $4$-scheme on $n>8$ points. 
Then there is a color $P\in\Part_2$ and its $\pi_3^3$-fiber $Q\in\Part_3$ such that 
$\pi_2^3(Q)=$ $\pi_3^3(Q)=P$ and the subdegree of $Q$ over $P$ is less than $\frac{n}{8}$.
\end{lemma}
\begin{proof}
Clearly, $\Part_1$ just has one color, say, $[n]$.
If $\Part_2$ has more than two colors then by antisymmetry it has at least $4$ colors and 
hence one of the colors $P\in\Part_2$ will have subdegree over $[n]$ less than $\frac{n}{4}$.
Again by the antisymmetry a $\pi_3^3$-fiber $Q\in\Part_3$ of $P$ will have subdegree 
$<\frac{n}{8}$ and $\pi_2^3(Q)=$ $\pi_3^3(Q)=P$.  

In the case when $\Part_2$ has just two colors - $P$ and its ``flipped'' color $P^T$ -
let us define: 
\begin{align*}
Q_1 &:= \{x\in[n]\mid (1,x)\in P\}\\
Q_2 &:= \{x\in[n]\mid (1,x)\in P^T\}\\
\end{align*}
Then obviously $Q_1, Q_2$ are disjoint sets of size $n_1:=\frac{n-1}{2}$ partitioning 
$\{2,\ldots,n\}$. Clearly, the image of the colors in $\Part_3$ restricting the first 
coordinate to $1$ gives us an antisymmetric partition $\Gamma$ of the sets $Q_1^{(2)}$, 
$Q_1\times Q_2$, $Q_2\times Q_1$ and $Q_2^{(2)}$; which is an association scheme on $Q_1^{(2)}$ 
and $Q_2^{(2)}$. By the antisymmetricity of $\Pi$, the colors 
corresponding to $Q_2\times Q_1$ are just the transpose (i.e. swap the two coordinates) 
of those corresponding to $Q_1\times Q_2$. Each color in $\Gamma$ can be naturally viewed 
as a $n_1\times n_1$ 
zero/one matrix. For example, a color $R$ corresponding to $Q_1\times Q_2$ can be represented 
as a matrix whose rows are indexed by $Q_1$ and whose columns are indexed by $Q_2$ such that: 
for all $(i,j)\in Q_1\times Q_2$, $R_{i,j}=1$ if $(i,j)\in R$ and $R_{i,j}=0$ if $(i,j)\not\in R$. 
Interestingly, in the matrix representation the composition property of Lemma 
\ref{lem-3-sch-assoc-sch} simply means that the linear combinations of the identity matrix $I$ and the 
colors in the partition of $Q_1\times Q_1$ (or $Q_2\times Q_2$) by $\Gamma$ is a matrix algebra, 
say $\Aalg_1$ (or $\Aalg_2$).

If $Q_1^{(2)}$ (or $Q_2^{(2)}$) is partitioned by $\Gamma$ into more than two parts then by 
antisymmetry there will be $\ge4$ parts which means that one of the parts will have subdegree 
$<\frac{n}{8}$. This gives us a required $\pi_3^3$-fiber $Q\in\Part_3$ of a $P\in\Part_2$. 

So we can assume that $Q_1^{(2)}$ and $Q_2^{(2)}$ are both partitioned into exactly two 
parts. Say,
\begin{itemize}
\item $R$ and $R^T$ are the two matrices representing the partition of $Q_1^{(2)}$ by $\Gamma$.
\item $S$ and $S^T$ are the two matrices representing the partition of $Q_2^{(2)}$ by $\Gamma$.
\end{itemize}
Note that: $R+R^T=S+S^T=J-I$ where $I$ is the identity matrix and $J$ is the all one matrix of
suitable dimensions. 

How do the partitions of $Q_1\times Q_2$ look like? Let $U$ be a matrix in the partition of
$Q_1\times Q_2$ by $\Gamma$. If $U=J$ (i.e. $\Gamma$ partitions $Q_1\times Q_2$ in a trivial way)
then by antisymmetricity $\Part_3$ has exactly $3!=6$ colors each of cardinality $n\cdot\#U=$ 
$n\cdot n_1^2$. But this is a contradiction as $6\cdot n\cdot n_1^2$ is not 
$n(n-1)(n-2)$. Thus,
$\Gamma$ partitions $Q_1\times Q_2$ into at least $2$ colors. Now since by antisymmetricity the 
number of colors in $\Part_3$ has to be a multiple of $6$, we deduce that $\Gamma$ partitions 
$Q_1\times Q_2$ into at least $4$ colors, say, $\{U_1,\ldots,U_4\}$. By the composition 
property of $\Gamma$, $U_1U_1^T$ is in $\Aalg_1$. In other words, there are positive integers 
$\alpha, \beta$ such that:
\begin{align*}
U_1U_1^T &= \alpha I + \beta (R+R^T)\\
\qquad &= \beta J+ (\alpha-\beta)I
\end{align*}
Thus, if $U_1$ is a singular matrix then $U_1U_1^T=\beta J$ implying that $U_1$ has equal rows. 
We can repeat the same argument with $U_1^T U_1$ (which is in $\Aalg_2$) and deduce that $U_1$ 
has equal columns. Now a zero/one matrix $U_1$ can have equal rows and equal columns iff $U_1=J$. 
This contradiction implies that $U_1$ is an invertible matrix. But then:
$$\{U_1U_1^T, U_1U_2^T, U_1U_3^T, U_1U_4^T\}$$
is a set of $4$ linearly independent matrices in $\Aalg_1$ which contradicts the fact that 
$\Aalg_1$ is a matrix algebra of dimension $3$. This contradiction implies that one of 
$Q_1^{(2)}$ or $Q_2^{(2)}$ is partitioned into at least four parts.

Thus, in all the cases the lemma is true.
\end{proof}

{}From the above lemma we see that at $2$ levels higher we get a suitable color with subdegree
reduced to a fraction of $2^{-3}$. This immediately gives us the following constant-factor 
improvement to Lemma \ref{lem-matching-in-logn}.

\begin{proposition}
If the $m$-scheme $\Pi:=\{\Part_1,\ldots,\Part_m\}$ on $n$ points is antisymmetric at the
first three levels, $|\Part_1|<n$ and $m\ge \frac{2}{3}\log_2 n$ then there is a matching
in $\{\Part_1,\ldots,\Part_m\}$.
\end{proposition}

\section{Primitivity of $m$-schemes and further research}

A $2$-scheme $\Pi=(\Part_1,\Part_2)$ on $n$ points can be viewed as a complete 
directed colored graph on $n$ vertices, where vertices of one color 
correspond to a $P\in\Part_1$ and the edges of one color correspond to a $Q\in\Part_2$. 
If an $m$-scheme is coming from a polynomial $f(x)$, over $k$, then we can try to relate graph 
properties of the $m$-scheme to the algebraic properties of the ideals defining the $m$-scheme. 
It turns out that such $m$-schemes can be efficiently
tested for one such property: connectivity. One can introduce a related 
notion: primitivity which is actually an extension of the primitivity of 
association schemes. 

Let $\Pi$ be a homogeneous $2$-scheme on the points $[n]$ 
with $\Part_2=\{P_{2,1},\ldots,P_{2,t_2}\}$. For every index
$i\in \{1,\ldots,t_2\}$ let $G_{2,i}$ denote the undirected 
graph on $[n]$ whose edges are unordered pairs $\{u,v\}$
where either $(u,v)\in P_{2,i}$ or $(v,u)\in P_{2,i}$. We say
that $\Pi$ is {\em primitive} if all the graphs $G_{2,1},\ldots,G_{2,t_2}$
are connected.

Let $I_{2,i}:=I^{\perp}(P_{2,i})$ be the ideal of
$\Aalg^{(2)}$ corresponding to $P_{2,i}$.
We define a subset $S(I_{2,i})$ of $\Aalg^{(1)}$ 
whose meaning would be clear later:
$$S(I_{2,i}):= \{ h\in\Aalg^{(1)}\ \mid\ (h\otimes 1-1\otimes h)\in I_{2,i}^\perp \}$$
It is easy to see that $k\subseteq S(I_{2,i})$ is a subalgebra of $\Aalg^{(1)}$. 
The following lemma relates the subalgebras $S(I_{2,i})$ to the
notion of primitivity.
\begin{lemma}
The dimension of the algebra $S(I_{2,i})$ over $k$ is equal to the number of
the connected components of the graph $G_{2,i}$. 
\end{lemma}
\begin{proof}
Let $G_{2,i}$ have $c$ connected components.
Observe that $h(x)\in S(I_{2,i})$ iff $(h(x_1)-h(x_2))I_{2,i}=0$ iff $h(u)=h(v)$ for all
$(u,v)\in\Supp(I_{2,i})$. The last condition precisely means that $h(x)$ is constant on the 
connected components of $G_{2,i}$. It follows that the polynomials $h_j(x)$, for $j\in[c]$, 
that are $1$ on all the vertices in the $j$-th connected component and $0$ on the rest, form 
a basis of $S(I_{2,i})$. Thus, the dimension of $S(I_{2,i})$ is $c$.
\end{proof}

The above lemma shows that if for some $i$ the graph $G_{2,i}$ is not connected (say, it has $c$ 
connected components) then (by solving a system of linear equations)
we compute a nontrivial subalgebra 
$S(I_{2,i})$ of $\Aalg^{(1)}$. This in explicit terms means that if $\Pi$ was obtained from a 
polynomial $f(x)$ of degree $n$ then we can compute $g(y)$ of degree $c$ such that 
$S(I_{2,i})\cong$ $k[y]/(g(y))$ and:
$$\Aalg^{(1)} \cong (k[y]/(g(y)))[x]/(\tilde{f}(y,x))$$
where, the $\text{deg}_x$ of $\tilde{f}(y,x)$ is $\frac{n}{c}$. Thus, we get 
two polynomials $g(y)$ and $\tilde{f}(y,x)$ of degrees $c$ and $\frac{n}{c}$ respectively 
to factor (the latter over the algebra 
$S(I_{2,i})\cong$ $k[y]/(g(y))$ rather than over the base field $k$). 
If we succeed in finding a nontrivial factor of either of these 
polynomials then we can find a zero divisor in $\Aalg^{(1)}$ and
then a factor of $f(x)$ therefrom. In particular, if $c\leq \sqrt{n}$
then it seems to be worth proceeding with factoring $g(y)$. 

We can generalize the notion of primitivity to higher levels as well.
\begin{definition}
Let $\Gamma=(\Part_1,\ldots,\Part_m)$ be a $m$-scheme. 
For a $P\in\Part_s$ such that 
$\pi^s_s(P)=$ $\pi^s_{s-1}(P)=:$ 
$Q\in\Part_{s-1}$, we fix $(v_1,\ldots,v_{s-2})\in \pi^{s-1}_{s-1}(Q)$.
We define the graph $G(P,v_1,\ldots,v_{s-2})$ on the vertex set 
$\{v\in [n]:(v_1,\ldots,v_{s-2},v)\in Q\}$ with edges $\{u,v\}$
such that either $(v_1,\ldots,v_{s-2},u,v)\in P$ or
$(v_1,\ldots,v_{s-2},v,u)\in P$. It turns out that connectedness of
$G(P,v_1,\ldots,v_{s-2})$ is independent of the choice of the tuple
$(v_1,\ldots,v_{s-2})$. We say that $\Gamma$ is primitive at level
$s$ if for every $P\in \Part_s$ with $\pi^s_s(P)=\pi^s_{s-1}(P)$,
the graph $G(P,\ldots)$ is connected. 
We say that $\Gamma$ is {\em primitive} if it is primitive at all levels $2\le s\le m$.
\end{definition}

Put
$I_{s,i}:=\Compli(P)$, $I_{s-1,i^\prime}:=\Compli(Q)$,
$I_{s-2,i''}:=\Compli(\pi^{s-1}_{s-1}(Q))$
  and define:
$$S(I_{s,i}):=\{ h\in I_{s-1,i^\prime}\ \mid\ (\iota_s^s(h)-\iota_{s-1}^s(h))\in I_{s,i}^\perp \}$$
One can show that $S(I_{s,i})$ is a subalgebra of $I_{s-1,i^\prime}$
and the number of connected components of $G(P,\ldots)$ is
$\frac{\dim_k S(I_{s,i})}{\dim_k I_{s-2,i''}}$. Thus in case
of imprimitivity, we can compute a subalgebra "between" 
$I_{s-2,i''}$ and $I_{s-1,i'}$ by solving a system 
of linear equations. If 
$1<\frac{\dim_k S(I_{s,i})}{\dim_k I_{s-2,i''}}
\leq \sqrt \frac{\dim_k I_{s-1,i'}}{\dim_k I_{s-2,i''}}$,
it seems to be worth proceeding with decomposing 
the ideal $I_{s-1,i'}$ by finding a zero divisor in 
the subalgebra $S(I_{s,i})$.

We feel that primitivity imposes strong conditions on the parameters of an $m$-scheme
but we do not know how to exactly use primitivity or imprimitivity 
and leave that for future research.

\section*{Acknowledgements}

N.S. would like to thank Centrum voor Wiskunde en Informatica, Amsterdam for
the postdoc fellowship. G.I. and N.S. would like to acknowledge the hospitality of 
Hausdorff Research Institute for Mathematics, Bonn where this work was partially done. 
We would like to thank Eiichi Bannai, Lajos R\'onyai and Ronald de Wolf for several 
interesting discussions.

\end{document}